\documentclass[aps,prb,showpacs,superscriptaddress,twocolumn]{revtex4-1}
\usepackage{latexsym}
\usepackage{amssymb}
\usepackage{bm}
\usepackage{graphicx}
\usepackage{booktabs}
\begin{document}
\title{Seebeck effect in dilute
  two-dimensional electron systems: temperature
  dependencies of diffusion and phonon-drag thermoelectric
  powers}
\author{S. Y. Liu}
\email{liusy@mail.sjtu.edu.cn}
\affiliation{Department of Physics, Shanghai
  Jiaotong University, 800 Dongchuan Road,
  Shanghai 200240, China}
\author{X. L. Lei}
\affiliation{Department of Physics, Shanghai
  Jiaotong University, 800 Dongchuan Road,
  Shanghai 200240, China}
\author{ Norman J. M. Horing}
\affiliation{Department of Physics and Engineering
  Physics, Stevens Institute of Technology,
  Hoboken, New Jersey 07030, USA}
\begin{abstract}
Considering screeening of electron scattering
interactions in terms of the
finite-temperature STLS theory and solving the
linearized Boltzmann equation (with no appeal to a relaxation
time approximation),
we present a theoretical
analysis of the low-temperature Seebeck effect in 
two-dimensional semiconductors with dilute
electron densities. We find that the temperature
($T$) dependencies of the diffusion and phonon-drag thermoelectric powers ($S_d$ and
$S_g$) can no longer be
described by the conventional simple
power-laws. As temperature increases, $|S_d|/T$
decreases when $T\gtrsim 0.1 \varepsilon_F$
($\varepsilon_F$ is the Fermi energy), while
$|S_g|$ first increases and then falls, resulting a peak
located at a temperature between Bloch-Gr\"uneisen temperature and $\varepsilon_F$.
\end{abstract}


\maketitle

\section{Introduction}
Heat generation with increasing density of
integrated electronic devices is one of
the serious restrictions blocking the
further development of conventional electronics.
To overcome this obstacle, there have been
proposals to use heat to storage
and transport information. Such an intriguing
concept provides hope of discovering new physics and
new green technology,  stimulating a great deal of
theoretical and experimental investigation. Recently,
two new subfields associated with
heat electronics have emerged, namely
"Phononics"\cite{Wang2008a} and "Spin Caloritronics".\cite{Bauer2010}
Phononics is devoted to the use of heat current to
perform computational operations and many heat devices,
such as heat diodes,\cite{Segal2008a,*Hu2006,*Li2005,*Segal2005,*Li2004,*Terraneo2002} heat
transistors\cite{Lo2008,*Segal2008,*Li2006}, thermal
memory\cite{Wang2008} {\it etc.}, have been proposed
and/or constructed. Spin Caloritronics concerns on
the motion of magnetization or of electron
spin induced by
heat or a temperature gradient.\cite{Bauer2010} In
this field, many phenomena, such as a spin Seebeck effect,\cite{Uchida2008,*Uchida2010,*Jaworski2010}
thermodynamic control of magnetization in 
 ferromagneto-nonmagnetic structures, 
\cite{[{For a review, see for
  example, }] Johnson2010} {\it etc.} have been reported.  

To use heat in electronics, the conventional
method is usually based on thermoelectric
effects, which convert a temperature gradient
to electric voltage. Among them, 
the Seebeck effect (SE), initially discovered
in metals by T. Seebeck in the
1820s,\cite{Seebeck1823,Seebeck1825} is widely used for
thermoelectric generation and for temperature
sensing. The first observations of SE
in bulk semiconductors, such as
Germanium\cite{Frederikse1953,Geballe1954} and
Silicon,\cite{Geballe1955} were reported in the 1950's. 
With the recent development of technology in
the fabrication of
semiconductor microstructures, investigations of
thermoelectric effects in two-dimensional electron
gases (2DEG) have been carried out both
experimentally\cite{OBLOH1984,Nicholas1984,Fletcher1985,Davidson1986,Fletcher1986,OBLOH1986,Ruf1988,Fletcher1988,Ying1994,Fletcher1994,Bayot1995,Tieke1996,Fletcher1997,Miele1998,Fletcher1998,Fletcher2002,Maximov2004,Zhang2004,Rafael2004,Chickering2009,Chickering2010}
and
theoretically.\cite{Girvin1982,ZAWADZKI1984,Nicholas1985,Cantrell1987,Cantrell1987a,Kubakaddi1987,Kundu1987,Lyo1988,Basak1989,Kubakaddi1989,Okuyama1990,VKKaravolasetal1990,VCKaravolasandPNButcher1991,XLLei1994,Zianni1994,Khveshchenko1996,Wu1996,Cooper1997,Khveshchenko1997,Tsaousidou1999,Karavolas2001,Kundu2001,Tsaousidou2001,Sankeshwar2005,Sergeev2005}

It is well known that, in thermoelectric power
(TEP), which is the main characteristic quantity of SE, there are two components; namely, diffusion thermopower,
$S_d$, and phonon-drag thermopower,
$S_g$.  At relatively low
 temperature, the diffusion process in the
 Seebeck effect has been expected to be dominant since 
the electron-phonon scattering is relatively weak.\cite{Nicholas1984,OBLOH1984,Fletcher1985,Davidson1986,OBLOH1986}
 However, a careful analysis of experimental data indicates
 that phonon-drag in 2DEGs also
 plays an important role even at temperature
 $T<10$\,K.\cite{Ruf1988,Fletcher1988,Fletcher1986,Wu1996,Okuyama1990,Kubakaddi1989,Lyo1988,Kundu1987,Kubakaddi1987,Cantrell1987a,Cantrell1987,Nicholas1985}
 Furthermore, there have also been studies of a
 sign change of diffusion TEP in a
 Si-MOSFET,\cite{VKKaravolasetal1990,VCKaravolasandPNButcher1991}
 and of the effects of weak localization on TEP,
\cite{Miele1998,Fletcher2002,Rafael2004} as well
as the TEP
 of composite-fermions,\cite{Chickering2010,Karavolas2001,Tsaousidou1999,Khveshchenko1997,Tieke1996,Bayot1995,Khveshchenko1996,Ying1994}
 and oscillation of TEP in low magnetic
 field,\cite{Fletcher1998,Zhang2004,Maximov2004}
 {\it etc.}

To understand the microscopic mechanisms in SE, it is
necessary to separate $S_d$ and
 $S_g$ from the total TEP that is measured.
 One way to do this is to analyze the temperature
 dependencies of $S_d$ and $S_g$. In the
absence of phonon-phonon scattering in the phonon
relaxation process, low-temperature $S_g$ {vs}
$T$ behavior has been taken to be of the form: $S_g\propto
\Lambda T^n$ with $\Lambda$ as the phonon mean
free path and $n=3$ or $n=4$ for dirty or clean
samples
respectively.\cite{Khveshchenko1997,Fletcher2002}
The diffusion TEP has often been
 assumed to vary linearly
with temperature.\cite{Fletcher1997} 
However, Sankeshwar, {\it
  et al.} showed that the inelastic
feature of electron-phonon scattering may result in
a nonlinear temperature
dependence of $S_d$ in
relatively clean 2D samples in the Bloch-Gr\"uneisen (BG)
regime, {\it i.e.} $T<T_{\rm BG}$ [the BG temperature $T_{\rm BG}\equiv
2k_Fu_{s\lambda}$ with $u_{s\lambda}$ as the
phonon velocity in branch
$\lambda$
 is about 5\,K for a 2DEG with typical
density $n_s\sim
10^{11}$\,cm$^{-2}$].\cite{Sankeshwar2005}
A few experiments were devoted to the direct
measurement of
diffusion TEP.\cite{Fletcher1994,Ying1994,Chickering2009}
Ying, {\it et al.} observed pure diffusion TEP
only in the case $T\le 0.2$\,K.\cite{Ying1994} Recently,
using hot electron thermocouple
structures, the diffusion TEP has been directly
detected by Chickering, {\it et al.} when $T<2$\,K.\cite{Chickering2009} 

It should be noted that the simple power-laws of $S_d$ and
$S_g$ vs $T$, obtained previously, were
derived on the basis of a relaxation
  time approximation (RTA), which is valid when
  $T, T_{\rm BG}\ll
  \varepsilon_F$. Recently, motivated by the observation
  of a so-called metal-insulator transition in
  resistivity vs temperature, clean undoped
  heterojunctions with electron density
  $n_s$ as low as $n_s\le 10^{10}$\,cm$^{-2}$ have
  been studied extensively.\cite{Zhou2010,Lilly2003,[{For a review, see for
  example, }] Spivak2010}  In these
  systems, $T$ and $T_{\rm BG}$ are comparable with
  the Fermi energy even at low temperature and therefore deviations
  of $S_d$ and $S_g$ {vs} $T$ from the
  conventional results are expected to be
  observed. 

In this paper, within the framework of Boltzmann
equation, we present a theoretical
investigation on thermoelectric effects in
2D electron GaAs/AlGaAs systems with carrier densities
$n_s=0.23\sim 1.06\times10^{10}$\,${\rm
  cm}^{-2}$. To account for the screening of
scattering interactions in a 2DEG with such low $n_s$, 
the finite-temperature
Singwi-Tosi-Land-Sjolander (STLS) theory, a scheme
beyond random
phase approximation (RPA), is
employed.\cite{Singwi1968,Schweng1994} Furthermore, to carefully treat inelastic
electron-phonon scattering, the Boltzmann equation
is solved with no appeal to a relaxation time approximation,
using an energy expansion method. 
Das Sarma and
Hwang have already presented a qualitative explanation of
experimental observations of resistivity in a 2DEG
with such dilute $n_s$ by means of a Boltzmann equation
combined with RPA-screened electron-impurity scattering.\cite{DasSarma2004,DasSarma2003,DasSarma1999}
In the present paper, performing numerical
calculations with STLS screening appropriate to
low carrier density, we find that
the temperature dependencies of $S_d$ and $S_g$ in
dilute 2D systems are significantly different from
those in the high-electron-density limit. When
  temperature increases, $|S_d|/T$ no longer remains
  unchanged:  it decreases for $T\gtrsim
  0.1\varepsilon_F$. In our calculation of phonon-drag TEP vs temperature, a
  peak appears: $|S_g|$ first increases and then
  falls as temperature increases.

 The paper is organized as follows. In Sec.\,II,
 an energy expansion method for solving the
 Boltzmann equation beyond the
RTA is presented along with the self-consistent finite-temperature STLS
theory.
Numerical investigation of the temperature
dependencies of diffusion and phonon-drag TEPs for
various dilute electron densities are exhibited in
Sec.\,III. Our results and conclusions are summarized in
Sec.\,IV. In the Appendix, we also provide
analytical results for $S_d$ and
$S_g$ {vs} $T$ in the high-electron-density limit, obtained by the energy
expansion method.

\section{Theoretical Considerations}

\subsection{Electron and phonon Boltzmann equations}

When a  two-dimensional electron with momentum
${\bm p}\equiv (p\cos\varphi_{\bm
  p},p\sin\varphi_{\bm p})$ and energy
$\varepsilon_{\bm p}=\frac{p^2}{2m^*}$ ($m^*$ is the effective electron mass) is subjected to a weak electric field $\bm E$
and a thermal gradient $\bm \nabla T$, its
kinetic motion can be described in terms of a nonequilibrium
distribution function, $f_{\bm p}$, which is
determined by a linearized Boltzmann equation of form 
\begin{equation}
\left (e{\bm E}\cdot {\bm v}_{\bm p}+\frac{\varepsilon_{\bm
      p}-\mu}{T}{\bm \nabla}T\cdot{\bm v}_{\bm p}\right
)\frac{\partial f_0(\varepsilon_{\bm p})}{\partial \varepsilon_{\bm p}}=I_{\rm scatt}.\label{BE1}
\end{equation}
Here, $\mu$ is the chemical potential, $T$ is the lattice temperature, $f_0(\varepsilon_{\bm p})=\{\exp[(\varepsilon_{\bm
  p}-\mu)/T]+1\}^{-1}$ is the equilibrium electron
distribution function and
${\bm v}_{\bm p}\equiv {\bm \nabla}_{\bm p}\varepsilon_{\bm p}={\bm
  p}/m^*$ is the electron velocity. In
Eq.\,(\ref{BE1}), $I_{\rm scatt}$ is the scattering
term due to electron-impurity and electron-phonon
interactions and it 
can be written as $I_{\rm scatt}=I_{\rm
    imp}+I_{\rm ph}$. $I_{\rm imp}$ represents
  the contribution to $I_{\rm scatt}$ from
  electron-impurity scattering:
\begin{equation}
I_{\rm imp}=-2\pi\sum_{\bm q}|\widetilde U_{\bm q}|^2\delta(\varepsilon_{\bm
  p}-\varepsilon_{{\bm p}-{\bm q}})\left(f_{\bm p} -f_{{\bm p}-{\bm q}}\right ),\label{SC_i}
\end{equation}
while $I_{\rm ph}$ is associated with the electron-phonon interaction:
\begin{eqnarray}
I_{\rm ph}&=&-{2\pi}\sum_{{\bm
    Q},\lambda,\pm}|\widetilde M_{{\bm Q
    }\lambda}|^2
\delta(\varepsilon_{{\bm
    p}\pm{\bm q}}-\varepsilon_{{\bm p}}\mp\Omega_{{\bm
    Q}\lambda})\nonumber\\
&&\times \left [N_{{\bm Q}\lambda}^{\pm}f_{\bm
    p}(1-f_{\bm p\pm \bm q})-N_{{\bm Q}\lambda}^{\mp}f_{\bm
    p\pm \bm q}(1-f_{\bm p})\right].\label{SC_p}
\end{eqnarray}
In Eqs.\,(\ref{SC_i}) and (\ref{SC_p}),
$\widetilde U_{\bm q}$ is the electron-impurity scattering potential,
$\widetilde M_{{\bm Q}\lambda}$ is the matrix element for interaction between
the 2D electrons and 3D phonons, and $N^\pm_{{\bm Q}\lambda}\equiv N_{{\bm
    Q}\lambda}+\frac{1}{2}\mp \frac{1}{2}$. $\Omega_{{\bm Q}\lambda}$ and
$N_{{\bm Q}\lambda}$, respectively, are the
energy and number of nonequilibrium phonons with
three-dimensional momentum ${\bm Q}\equiv ({\bm q},q_z)=(q_x,q_y,q_z)$ in branch
$\lambda$. 

 Since the temperature gradient may drive the phonons
 out of equilibrium, $N_{{\bm Q}\lambda}$ in Eq.\,(\ref{SC_p})
differs from the number of equilibrium phonons, $n_{{\bm
    Q}\lambda}\equiv \left [\exp(\Omega_{{\bm
      Q}\lambda}/T)-1\right ]^{-1}$, and it is determined by the
Boltzmann equation for phonons:
\begin{equation}
\frac{d N_{{\bm Q}\lambda}}{d t}=\left (\frac{\partial N_{{\bm
        Q}\lambda}}{\partial t}\right )_{\rm d}
+\left (\frac{\partial N_{{\bm
        Q}\lambda}}{\partial t}\right )_{\rm ep}
+\left (\frac{\partial N_{{\bm
        Q}\lambda}}{\partial t}\right )_{\rm bp}.
\end{equation}
Here, $\left(\frac{\partial N_{{\bm
        Q}\lambda}}{\partial t}\right)_{\rm d}$ is
the drift term, taking the form 
\begin{equation}
\left (\frac{\partial N_{{\bm
        Q}\lambda}}{\partial t}\right )_{\rm
  d}=-{\bm u}_{\bm Q\lambda}\cdot{\bm \nabla}N_{{\bm Q}\lambda},\label{BEP}
\end{equation}
with ${\bm u}_{\bm Q\lambda}$ as the phonon velocity.
Note that, in the present paper, the magnitudes of ${\bm
  u}_{\bm Q\lambda}$ are assumed to be
independent of $\bm Q$ and they are denoted by $u_{s\lambda}$ (longitudinal and
transverse acoustic phonons are denoted by $u_{sl}$ and
$u_{st}$, respectively).
$\left (\frac{\partial N_{{\bm
        Q}\lambda}}{\partial t}\right )_{\rm
  bp}$
 is the relaxation
term due to the boundary and phonon-phonon
scatterings, written as
\begin{equation}
\left (\frac{\partial N_{{\bm
        Q}\lambda}}{\partial t}\right )_{\rm
  bp}=-\left (\frac{1}{\tau_{\rm bs}}+\frac 1{\tau_{\rm pp}}\right)\left (N_{{\bm Q}\lambda}-n_{{\bm Q}\lambda}\right),
\end{equation}
with $\tau_{\rm bs}$ as the relaxation time due
to boundary scattering, $1/\tau_{\rm bs}=u_{\bm Q\lambda}/\Lambda$,\cite{Okuyama1990} and $\tau_{\rm
  pp}$ as the relaxation time due to
phonon-phonon scattering, $1/\tau_{\rm
  pp}=A_\lambda T^3\Omega_{\bm Q\lambda}^2$.\cite{Herring1954,Callaway}
$\left (\frac{\partial N_{{\bm
        Q}\lambda}}{\partial t}\right )_{\rm ep}$ is the phonon
scattering rate due to the
electron-phonon interaction, as given by
\begin{eqnarray}
\left (\frac{\partial N_{{\bm
        Q}\lambda}}{\partial t}\right )_{\rm
  ep}&=&-\frac{2\pi}{L}g_s|\widetilde M_{{\bm Q}\lambda}|^2\sum_{{\bm
    p},\pm}\left \{\pm\delta(\varepsilon_{\bm p}-\varepsilon_{{\bm p}\pm{\bm
      q}}\pm\Omega_{{\bm Q}\lambda})\right .\nonumber\\
&&\times \left . N^\pm_{{\bm Q}\lambda}f_{\bm
    p}(1-f_{{\bm p}\pm{\bm q}})\right \}, \label{SC_EP}
\end{eqnarray}
with $g_s$ as the spin degeneracy and $L$ as the
sample size along
the direction perpendicular to the 2D sheet. 
In the case of a weak temperature gradient, Eq.\,(\ref{BEP}) can be
solved analytically and
the steady-state number of nonequilibrium phonons can be written as
\begin{equation}
N_{{\bm Q}\lambda}=n_{{\bm Q}\lambda}-\tau_{{\rm
    p}\lambda}{\bm u}_{\bm Q\lambda}\cdot
{\bm \nabla} T\frac{\partial n_{{\bm Q}\lambda}}{\partial T}.\label{NEQ}
\end{equation}
Here, $1/\tau_{{\rm p}\lambda}\equiv 1/\tau_{{\rm
    bs}}+1/\tau_{{\rm pp}}+1/\tau_{\rm ep}
$ and $\tau_{\rm ep}$ is the phonon relaxation time
due to electron-phonon
scattering, taking the form
\begin{equation}
\frac{1}{\tau_{\rm ep}}=\frac{2\pi g_s}{L}\sum_{\bm
  p}|\widetilde M_{{\bm Q}\lambda}|^2\delta(\varepsilon_{{\bm p}-{\bm q}}-\varepsilon_{\bm
  p}+\Omega_{{\bm Q}\lambda})[f_0(\varepsilon_{{\bm p}-{\bm
    q}})-f_0(\varepsilon_{\bm p})].
\end{equation}

\subsection{Energy-expansion method to solve
  electron Boltzmann equation}

To solve the electron Boltzmann equation,
Eq.\,(\ref{BE1}), we
assume that the nonequilibrium distribution function $f_{\bm
  p}$ takes the form 
\begin{equation}
f_{\bm p}=f_0(\varepsilon_{\bm p})+g_{\bm p}\left [-\frac{\partial f_{0}(\varepsilon_{\bm
      p})}{\partial \varepsilon_{\bm p}}\right ],
\end{equation}
with $g_{\bm p}$ as an unknown function. In previous studies,  
when electron-optical-phonon scattering can be ignored at low
temperature, $g_{\bm
  p}$ is usually obtained using the relaxation time
approximation (RTA). Obviously, RTA is valid only in
the high-electron-density limit.
In the present paper,  in order to study diffusion
and phonon-drag TEPs for
relatively low electron density, we follow the idea proposed by
Allen for an investigation of transport in
metals,\cite{Allen1978} which assumes that 
 $g_{\bm p}$ can be expanded in terms of basis
 functions $\chi_{Jn}({\bm p})$:
\begin{equation}
g_{\bm p}=\sum_{J,n} C_{Jn}\chi_{Jn}({\bm p}),\label{Expand}
\end{equation}
with $C_{Jn}$ as the coefficients of expansion. In a 2D system with
a parabolic dispersion relation, the functions
$\chi_{Jn}({\bm p})$ can be written as
\begin{equation}
\chi_{Jn}(\bm p)=4\pi F_J({\bm p})\frac{\eta_n(\varepsilon_{\bm p})}{p}.
\end{equation}
 Here, $F_J(\bm p)$ are the basis functions for
the expansion of $g_{\bm p}$ with respect to
 the angle of the momentum vector $\bm p$, and they can be
 chosen as sine or cosine
 functions of multiples of the angle $\varphi_{\bm p}$. $\eta_n(\varepsilon)$ are $n$-$th$
order polynomials in electron energy $\varepsilon$
and they are 
orthogonal with respect to the weight function $-\partial
f_0(\varepsilon)/\partial \varepsilon$:
\begin{equation}
\int_0^\infty \left (-\frac{\partial
    f_0(\varepsilon)}{\partial \varepsilon}\right
)\eta_n(\varepsilon)\eta_{m}(\varepsilon)d\varepsilon=\delta_{nm}.\label{OR}
\end{equation}
It is noted that to study the transport in metals, the lower limit of
energy integration in Eq.\,(\ref{OR}) can be assumed to be $-\infty$, since the Fermi
energy in metals usually is much larger than the bottom of electron energy band.\cite{Allen1978} However, in three- or two-dimensional
semiconductors, the finite bottom of the energy band or subband may
affect transport properties, especially at relatively high
temperature (or for relatively low electron density). Hence, in Eq.\,(\ref{OR}), the lower limit of integration
is maintained equal to zero. Further, in our study, we assume that
$\eta_n(\varepsilon)$ take the form
\begin{equation}
\eta_n(\varepsilon)=\sum_{m=0}^n\alpha_{nm}\varepsilon^m,\label{eta_n}
\end{equation}
which also differs from that proposed by
Allen.\cite{Allen1978} In Eq.\,(\ref{eta_n}), the parameters
$\alpha_{nm}$ are determined from the
orthonormality conditions of
$\eta_n(\varepsilon)$. In general,
they are independent of $\varepsilon$ but
may depend on the lattice temperature, as well as
on the Fermi energy $\mu$. Note that for
$n=0$, $\eta_0(\varepsilon)$ is an
energy-independent constant: $\eta_0(\varepsilon)=\eta_0$.     

Furthermore, without loss of generality, we assume that the electric
field and temperature gradient are applied along
the $x$ axis. Thus, in 2D semiconductors with
parabolic dispersion, only one term with basis
function $F_{J=X}({\bm
  p})=\cos\varphi_{\bm p}$ need be considered in
the expansion of $g_{\bm p}$ with respect to
$\varphi_{\bm p}$. Multiplying both sides of
Eq.\,(\ref{BE1}) by $\chi_{n}({\bm p})$
[$\chi_{n}({\bm p})\equiv \chi_{n,J=X}({\bm
  p})=4\pi\cos\varphi_{\bm
  p}\eta_n(\varepsilon_{\bm p})/p$] and performing
the summation
over ${\bm p}$, the linearized Boltzmann equation
for electrons can be rewritten as 
\begin{equation}
-\frac{eE}{\eta_0}\delta_{n0}-\frac{\nabla_x
  T}T\sum_{m=0}^n\alpha_{nm}\gamma_m+\frac{\nabla_x T}{T}D_n=\sum_{n'=0}^{\infty}Q_{nn'}C_{n'}\label{BE_R}
\end{equation} 
with $\gamma_{m}=\int_0^\infty d\varepsilon (\varepsilon-\mu)\varepsilon^{m} 
\frac{\partial f_0(\varepsilon_{\bm
    p})}{\partial \varepsilon_{\bm p}}$. In this
equation, the third term on left-hand side
is the source of the phonon-drag
effect: it describes the interaction
between equilibrium electrons and nonequilibrium
phonons. In it, $D_n$ take the form
\begin{eqnarray}
D_n&=&2\pi\sum_{{\bm Q},{\bm
    p},\lambda,\pm}(\mp 1)|\widetilde M_{{\bm Q
    }\lambda}|^2\chi_{n}(\varepsilon_{\bm p})
\delta(\varepsilon_{{\bm
    p}-{\bm q}}-\varepsilon_{{\bm p}}\mp\Omega_{{\bm
    Q}\lambda})\tau_{{\rm p}\lambda} \nonumber\\
&&\times u^x_{\bm
  Q\lambda}\frac{\Omega_{\bm Q
    \lambda}}{T}n_{{\bm Q}\lambda}(1+n_{{\bm
    Q}\lambda})\left [f_0(\varepsilon_{\bm p})-f_0(\varepsilon_{{\bm p}-{\bm
      q}})\right ],\label{D}
\end{eqnarray}
with $u^x_{\bm Q\lambda}$ as the $x$ component of
${\bm u}_{\bm Q\lambda}$.
Note that Eq.\,(\ref{D}) is derived from
Eq.\,(\ref{SC_p}) by substituting into it the explicit
form of the number of nonequilibrium phonons, {\it
  i.e.} Eq.\,(\ref{NEQ}).
On right-hand side (r.h.s.) of Eq.\,(\ref{BE_R}), $Q_{nn'}$ are 
associated with the scattering term $I_{\rm sc}$
and they can be written
as $Q_{nn'}=Q_{nn'}^{\rm imp}+Q_{nn'}^{\rm ph}$
with $Q_{nn'}^{\rm imp}$ and $Q_{nn'}^{\rm ph}$,
respectively, taking the forms ($\varphi_{\hat{\bm p\bm
  q}}$ is the angle between $\bm p$ and $\bm q$)
\begin{eqnarray}
Q_{nn'}^{\rm imp}&=&16\pi^3\sum_{\bm p,\bm
  q}|\widetilde U_{\bm q}|^2\left [-\frac{\partial
    f_0(\varepsilon_{\bm p})}{\partial
    \varepsilon_{\bm p}}\right ]\frac{q}{p^3}\eta_n(\varepsilon_{\bm p})\eta_{n'}(\varepsilon_{\bm p})\nonumber\\
&&\times\delta(\varepsilon_{\bm p}-\varepsilon_{{\bm p}-{\bm q}})\cos\varphi_{\hat{\bm p\bm q}}
\end{eqnarray}
and
\begin{eqnarray}
Q_{nn'}^{\rm ph}&=&\frac{8\pi^3}{m^*T}\sum_{\bm
  p,\bm Q,\lambda,\pm}|\tilde M_{\bm Q,\lambda}|^2\delta(\varepsilon_{\bm p-\bm q}-\varepsilon_{\bm p}\mp\Omega_{\bm Q,\lambda})n_{\bm Q\lambda}^\pm\nonumber\\
&&\times f_0(\varepsilon_{\bm p})[1-f_0(\varepsilon_{\bm p-\bm q})]\frac{1}{\sqrt{\varepsilon_{\bm p}}}\eta_n(\varepsilon_{\bm p})\nonumber\\
&&\times\left [\frac{\eta_{n'}(\varepsilon_{\bm p})}{\sqrt{\varepsilon_{\bm p}}}-\frac{\eta_{n'}(\varepsilon_{\bm p-\bm q})}{\varepsilon_{\bm p-\bm q}\sqrt{2m^*}}(p-q\cos\varphi_{\hat{\bm p\bm q}})\right ].\label{Qph}
\end{eqnarray}

Thus, the original linearized Boltzmann equation
is reduced to Eq.\,(\ref{BE_R}), a system of linear
equations for $C_n$. 
After $C_{n}$ are determined, the macroscopic charge
current can be evaluated through
\begin{equation}
{J}_{x}=-g_se\sum_{\bm p}\left [-\frac{\partial
  f_0(\varepsilon_{\bm p})}{\partial
  \varepsilon_{\bm p}}\right ]g_{\bm p}{v}_{\bm
px}=-g_se C_{0}/\eta_0^2.\label{Jx}
\end{equation} 
Since there are three driving terms in
Eq.\,(\ref{BE_R}), its solution, $C_0$,
can be written as $C_0=C_0^{(\rm c)}+C_0^{(\rm
  d)}+C_0^{(\rm g)}$ with $C_0^{(\rm c)}$, $C_0^{(\rm
  d)}$, and $C_0^{(\rm g)}$ determined from
Eq.\,(\ref{BE_R}) in the presence of only the
first, the second, or the third driving term,
respectively. Obviously, $C_0^{(\rm c)}$ is
proportional to $E$ and it determines the
conductivity as $\sigma=-g_seC_0^{(\rm c)}/(\eta_0^2 E)$. $C_0^{(\rm
  d)}$ and $C_0^{(\rm g)}$ are proportional to
$\nabla_x T$ and they are associated with the diffusion and
phonon-drag TEPs, respectively:
$S_{d}=-g_seC_0^{(\rm d)}
/(\eta_0^2\nabla_xT\sigma)$ and
$S_{g}=-g_seC_0^{(\rm g)} /(\eta_0^2\nabla_xT\sigma)$.

We note that such an energy expansion method
presented here can also reproduce the previous RTA results
in high-electron-density limit. We present a detailed
calculation of the high-$n_s$ TEP as a function of
$T$ in the Appendix, considering screened
electron-impurity scattering as well as screened
piezoelectric interaction and unscreeened
deformation interaction between electron-acoustic
phonons. There, the well-known Mott relation is obtained
for $S_d$, and the lowest-order correction to the Mott formula at
low temperature may come not only from
electron-impurity scattering but also from the
interaction between electrons and acoustic phonons
when $T$ lies within the equipartition (EP) regime,
$T_{\rm BG}\ll T\ll \varepsilon_F$. We also
obtain the well-known $T^4$ law for $S_g$ {vs} temperature for
$T$ within the BG regime. Besides, in the presence
of only boundary scattering in the phonon relaxation
process, $S_g$ is found to be independent of
temperature when $T_{\rm BG}\ll T\ll
\varepsilon_F$.

\subsection{Finite-temperature STLS theory}

To analyze resistivity as a function of $T$ in
dilute 2D systems, it is necessary to clarify the
role of screening in
electron-impurity and electron-acoustic-phonon
scatterings. Using
RPA-screened electron-impurity scattering, Das Sarma and
Hwang have qualitatively explained the
experimental observations in a dilute
2DEG.\cite{DasSarma2004,DasSarma2003,DasSarma1999}
However, in the GaAs systems that we study, the
dimensionless Wigner-Seitz density (or
interaction) parameter $r_s=1/(a_B\sqrt{\pi n_s})$
[$a_B=4\pi\varepsilon_0\kappa /(m^*e^2)$ is
the effective semiconductor Bohr radius and
$\kappa$ is the background dielectric constant]
can reach the value $\sim 11.6$ for 2D GaAs with
$n_s=0.23\times 10^{-10}$\,cm$^{-2}$. Hence, the
local-field correction to RPA is quite important
and we therefore use the finite-temperature self-consistent
STLS theory here. 

Within the framework of the Boltzmann equation
approach with interaction screening included, the scattering
potential is usually divided by the dielectric function
 $\varepsilon(q,\omega)$, which takes the form
\[
\varepsilon(q,\omega)=1-V_qH(q)[1-G(q)]\chi_0(q,\omega).
\]
Here, $\chi_0(q,\omega)$ is the density-density
correlation function of the free 2D system, $H(q)$ is the form factor
of the electron-electron interaction in
the 2D system, and $V_q=e^2/(2\kappa\varepsilon_0q)$
is the 2D Coulomb potential. $G(q)$ is the static
local-field factor whose value depends on 
the approximation that used.  In RPA, $G(q)$ is
zero, while $G(q)=q/\left
  (2\sqrt{q^2+k_F^2}\right)$ in Hubbard's approximation.\cite{Jonson1976}
In STLS theory which we use here, the local field
factor is determined by the structure factor
$S(q)$ through
\begin{eqnarray}
G(q)&=&-\frac{1}{n_s}\sum_{\bm k}\frac{{\bm
    k}\cdot {\bm q}}{kq}\frac{H(k)}{H(q)}\left
  [S(|{\bm k}-{\bm q}|)-1\right ].\label{STLS_G}
\end{eqnarray} 
On the other hand, $S(q)$ is also related to $G(q)$ via
\begin{eqnarray}
S(q)=-\frac{T}{n_s}\sum_{n=-\infty}^{\infty}\chi(q,2i\pi nT),\label{STLS_S}
\end{eqnarray} 
with $\chi(q,i\omega)\equiv
\chi_0(q,i\omega)/\varepsilon(q,i\omega)$ as the
response function. Thus, Eqs.\,(\ref{STLS_G}) and
(\ref{STLS_S}) form a closed system of equations,
to be solved self-consistently by iteration.
 
\section{Results and Discussion}

We carry out numerical calculations to investigate the
thermoelectric effect of a dilute 2D electron gas in
a GaAs/AlGaAs heterojunction at temperature $T<5$\,K.
The electron Boltzmann equation is solved by means
of the energy expansion method and the screening of
scattering is evaluated self-consistently within
the framework of the finite-temperature STLS theory.
In these calculations, the screened
electron-impurity scatterings due to both
remote and background impurities are considered.
The corresponding scattering potential takes the
form\cite{Lei1985}
\begin{equation}
|\widetilde U_{\bm q}|^2=|U_{\bm
  q}|^2/|\varepsilon(q,0)|^2\label{POT}
\end{equation}
with $|U_{\bm q}|^2={V_q^2}\left
  [N_{\rm r}{\rm
  e}^{-2qs}I(q)^2+{N_{\rm b}}J(q)/q\right ]$,
$I(q)$ and $J(q)$ are the form factors, $N_{\rm b}$
is the density of background impurities, and
$N_{\rm r}$ represents the density of
remote impurities located at distance $s$ from the
heterojunction interface on the AlGaAs side. 

In regard to the electron-phonon interaction,
only acoustic phonons contribute to
scattering at low temperature. The corresponding
potential can be written as
\begin{equation}
|\widetilde M_{\bm Q\lambda}|^2=|M_{\bm Q\lambda}|^2|I(iq_z)|^2
\end{equation}
with $|M_{\bm Q\lambda}|^2$ as the matrix element
of the electron-phonon interaction in
three-dimensional plane-wave representation.
 In present paper, we consider both
the deformation and piezoelectric interactions
between electrons and acoustic phonons. It is well
known that only the longitudinal acoustic
phonon (LA) mode gives rise to deformation scattering with
matrix element
\begin{equation}
|M_{{\bm Q},{\rm
LA}}|_{\rm def}^2=\frac{{\Xi}^2Q}{2du_{sl}}.\label{DP_M}
\end{equation}
Here, $d$ is the mass density of
crystal and 
$\Xi$ is the shift of the band edge per unit
dilation. Both the longitudinal and
transverse (TA) acoustic phonons contribute to the
piezoelectric interaction. The corresponding
scattering matrix elements take the forms\cite{Lei1985}
\[
|M_{{\bm Q},{\rm
LA}}|_{\rm piez}^2=\frac{32\pi^2
e^2e_{14}^2}{\kappa^2du_{sl}|\varepsilon(q,\Omega_{\bm Q,{\rm LA}})|^2}\frac{9q_x^2q_y^2q_z^2}{Q^7},
\]
and
\begin{eqnarray}
|M_{{\bm Q},{\rm
TA}}|_{\rm piez}^2&=&\frac{32\pi^2
e^2e_{14}^2}{\kappa^2du_{st}Q^5|\varepsilon(q,\Omega_{\bm Q,{\rm TA}})|^2}\nonumber\\
&&\times\left
(q_x^2q_y^2+q_x^2q_z^2+q_y^2q_z^2-\frac{9q_x^2q_y^2q_z^2}{Q^2}\right ),\label{PZ_M}
\end{eqnarray}
with $e_{14}$ as the piezoelectric constant.

In Eqs.\,(\ref{DP_M}) and (\ref{PZ_M}), the
unscreened form of the electron-acoustic-phonon
scattering through deformation potential is
used, while the piezoelectric interaction is
assumed to be dynamically screened. Such a treatment is based on
the fact that these two interactions have
completely different origins. It is well known
that piezoelectric
electron-phonon scattering comes from the Coulomb interaction
of electrons in an electric field
induced by thermal vibration of atoms, and hence it is
effectively screened by
electron-electron interactions. However, the
deformation scattering mainly
results from the overlap of electron wave
functions between different atoms in distorted lattices.\cite{Bardeen1950} Thus, the deformation interaction between electrons and phonons does
not directly relate to the Coulomb
interaction, and therefore it is inappropriate to use the
screened form for the deformation potential.
We note that, employing the unscreened form for the deformation
 potential with the appropriate
parameter $\Xi$, good
agreement between theory and
 experiments has been reached in a previous study
on phonon-drag thermoelectric effect.\cite{Okuyama1990}

In our numerical calculations, the parameters are chosen as follows:
$\kappa=12.9$, $d=5.31$\,g/cm$^{3}$,
$u_{ sl}=5.29\times 10^3$\,m/s, $u_{st}=2.48\times 10^3$\,m/s,
$\Xi=8.5$\,eV, $m^*=0.067m_0$ ($m_0$ is free
electron mass), $e_{14}=1.41\times 10^9$\,V/m. 
Since we are interested in
the temperature and electron-density dependencies
of the diffusion and phonon-drag TEPs
at low temperature ($T\le 5$\,K),
the relaxation of phonons due to phonon-phonon
scattering can be ignored and only the
temperature-independent boundary scattering need be
considered. Furthermore,
the phonon mean free path is assumed to be
$\Lambda=2.42$\,mm.\cite{Tsaousidou1999}
The truncation of summation in the expansion of $g_n$
is estimated by the convergence of the numerical scheme.
We find that, for $n_s\ge 0.2\times
10^{10}$\,cm$^{-2}$ and $T\le 5$\,K,  $n_{\rm max}=4$ 
is sufficient to reach the required numerical accuracy. 

The low-temperature transport properties
depend sensitively on the impurity densities. In
the present paper, to obtain results in qualitative
agreement with the experimental resistivity data of
Ref.\,\onlinecite{Lilly2003}, the density of
charges in the depletion
layer is chosen to be $N_{\rm dep}=7\times
10^{9}$\,cm$^{-2}$ and the background
impurity density is assumed to be constant: $N_{\rm b}=1\times
10^{18}$\,m$^{-3}$. The remote impurity density,
$N_{\rm r}$, is determined from the 
mobility at $T=30$\,mK by assuming $s=210$\,nm. In 
Fig.\,1, we plot the dependencies of resistivity
$\rho=1/\sigma$ on temperature for various
electron densities. An evident ``metal-insulator''
transition can be observed: when $T$ increases
$\rho$ increases for dense $n_s$, while it
decreases for dilute $n_s$. Such behavior
of $\rho$ vs $T$ almost agrees quantitatively with
experimental data in the case $1\,{\rm K}\le T\le 5\,{\rm
  K}$ for all $n_s$ which were studied (see Fig.\,2 in
Ref.\,\onlinecite{Lilly2003}). However, in Fig.\,1,
we do not see the small peaks for
intermediate $n_s$, which have been observed
experimentally.\cite{Lilly2003} This is associated
with the fact that the observed
small peaks in $\rho$ vs $T$ are the result of weak (or strong) localization,
which is ignored in our study. 

\begin{figure}
\includegraphics [width=0.45\textwidth,clip] {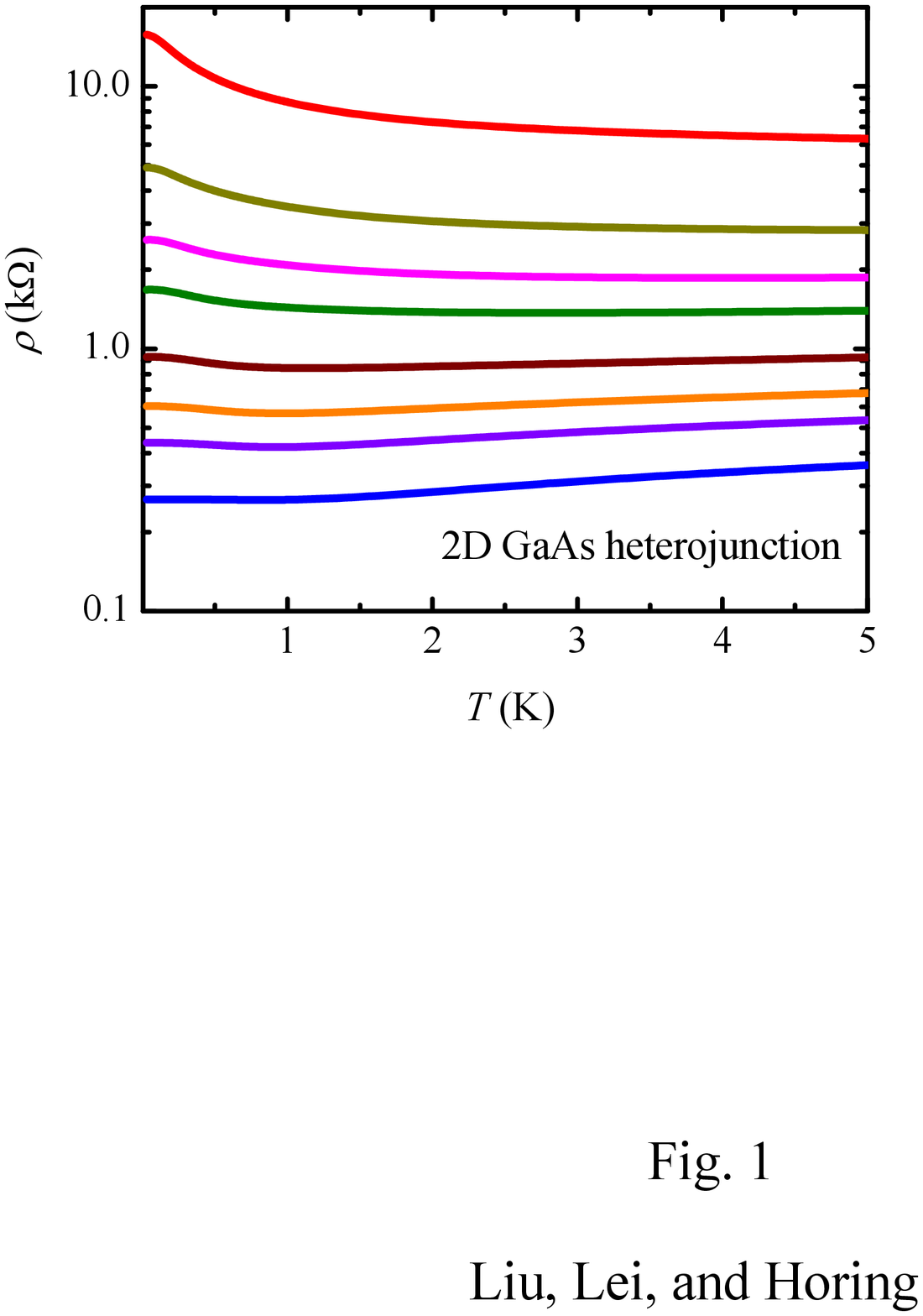}
\caption{(Color online) Temperature dependencies of resistivity
for various densities of electrons (from the top): $n_s=0.23$,
  $0.29$, $0.36$, $0.42$, $0.55$, $0.68$,
  $0.80$, and $1.06\times
  10^{10}$\,cm$^{-2}$. The remote impurity sheet
is assumed to be located at
  $210$\,nm from the interface on AlGaAs
side. The density of charges in the depletion layer is
assumed to be $N_{\rm dep}=7\times
10^9$\,cm$^{-2}$ and a constant  density of
background impurities is used: $N_{\rm b}=1\times 10^{18}$\,m$^{-3}$. } \label{fig1}
\end{figure}

\begin{figure}
\includegraphics [width=0.45\textwidth,clip] {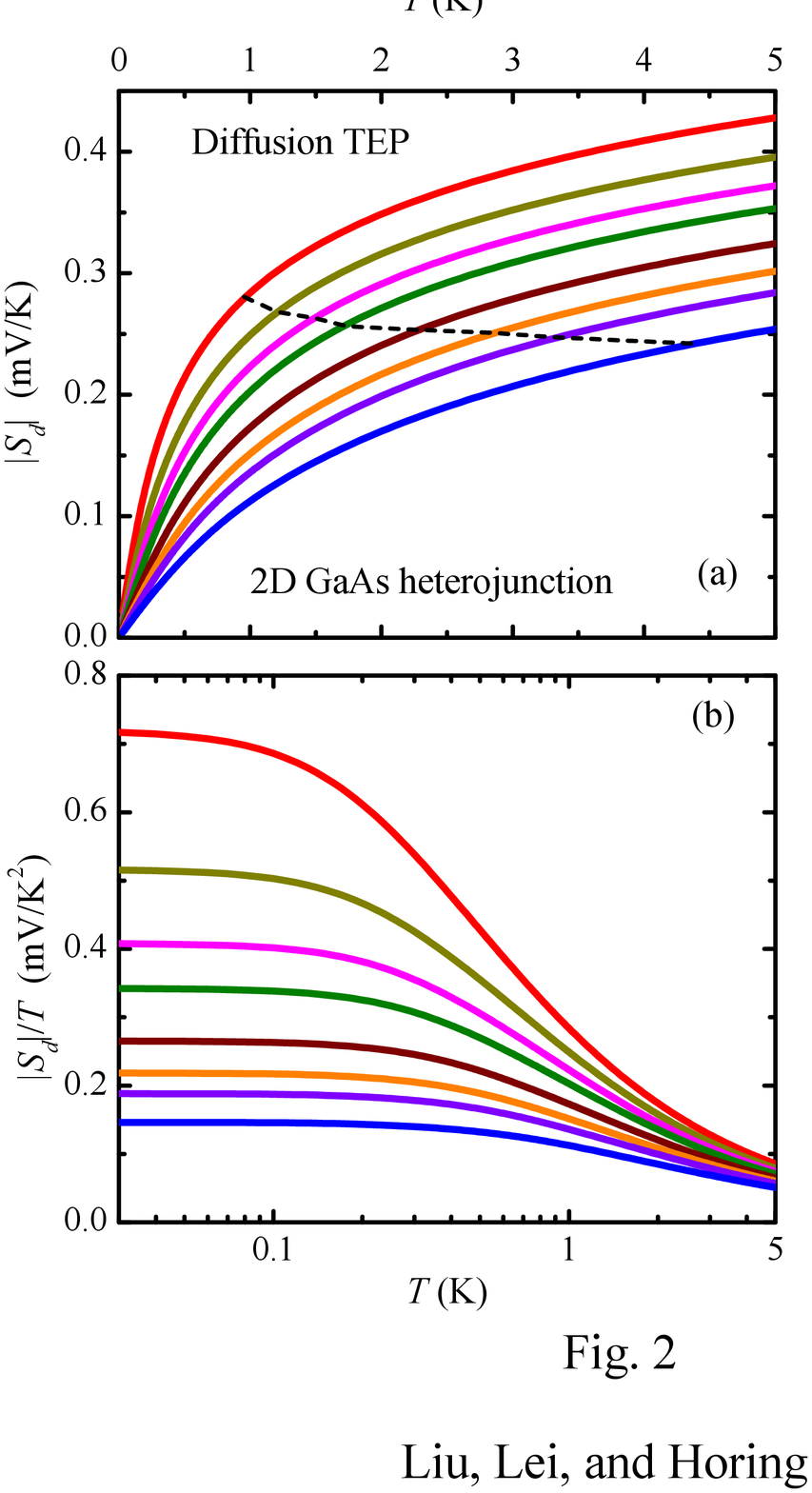}
\caption{(Color online) Temperature dependencies of (a) $|S_d|$ and (b) $|S_d|/T$
  for various densities of electrons (from the top): $n_s=0.23$,
  $0.29$, $0.36$, $0.42$, $0.55$, $0.68$,
  $0.80$, and $1.06\times
  10^{10}$\,cm$^{-2}$. The dashed line in
  Fig.\,2(a) indicates the Fermi energy
  $\varepsilon_F$ (in Kelvin).} \label{fig2}
\end{figure}

In Fig. 2, we plot the temperature dependencies of $|S_d|$ and
$|S_d|/T$ in a 2D GaAs heterojunction for various
electron densities in the range $0.23\le n_s\le 1.06\times 10^{10}$\,cm$^{-2}$. From Fig.\,2(a), we see that,
with an increase of temperature, $|S_d|$
increases. However, this increase is no
longer linear. To clearly show the nonlinear
dependencies of $S_d$ on $T$,
the temperature dependencies
of $|S_d|/T$ for $T\le 5$\,K are plotted in Fig.\,2(b). We see that
when temperature increases, $|S_d|/T$ remains
constant only for $T\lesssim 0.1 \varepsilon_F$.
Beyond this regime,  $|S_d|/T$ decreases with
an increase of temperature. Such nonlinear dependence of $S_d$
on temperature mainly stems from broadening of
the electron distribution function at relatively high
temperature. 

It should be noted that electron-phonon scattering
also may affect $|S_d|$ at relatively high
temperature. To show this, in Fig.\,3, 
$S_d$ vs $T$ is plotted both in the absence and in the presence of
electron-phonon interactions. It is
clear that the contribution from electron-phonon
scattering to $S_d$ is important for relatively
high $n_s$. This is associated with the fact that
for dilute $n_s$, electron-impurity scattering is
so strong that the electron-phonon interaction is
relatively unimportant within the temperature regime
studied. From Fig.\,3, we also see
that the magnitude of $S_d$ in the presence of
electron-phonon interaction is always less than
that in the absence of electron-phonon scattering,
reflecting the fact that
contribution to $|S_d|$ from electron-phonon
scattering is negative.
   
\begin{figure}
\includegraphics [width=0.45\textwidth,clip] {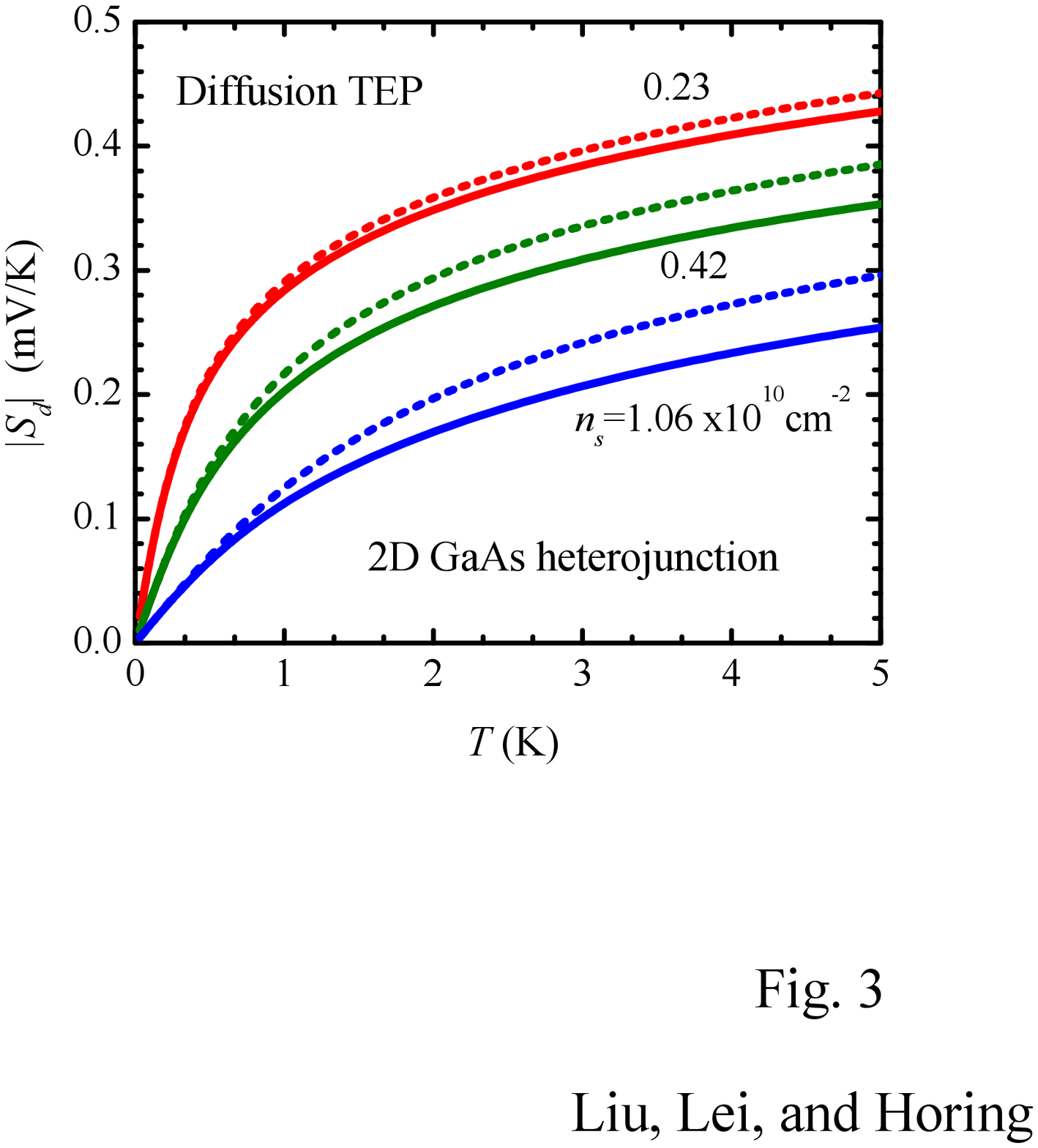}
\caption{(Color online) Effect of electron-phonon scattering on
  diffusion TEP for $n_s=0.23$,
  $0.42$, and $1.06\times
  10^{10}$\,cm$^{-2}$. The solid and dotted lines
  indicate $|S_d|$ in the presence and in the absence of
  electron-phonon scattering, respectively.} \label{fig3}
\end{figure}

\begin{figure}
\includegraphics [width=0.45\textwidth,clip] {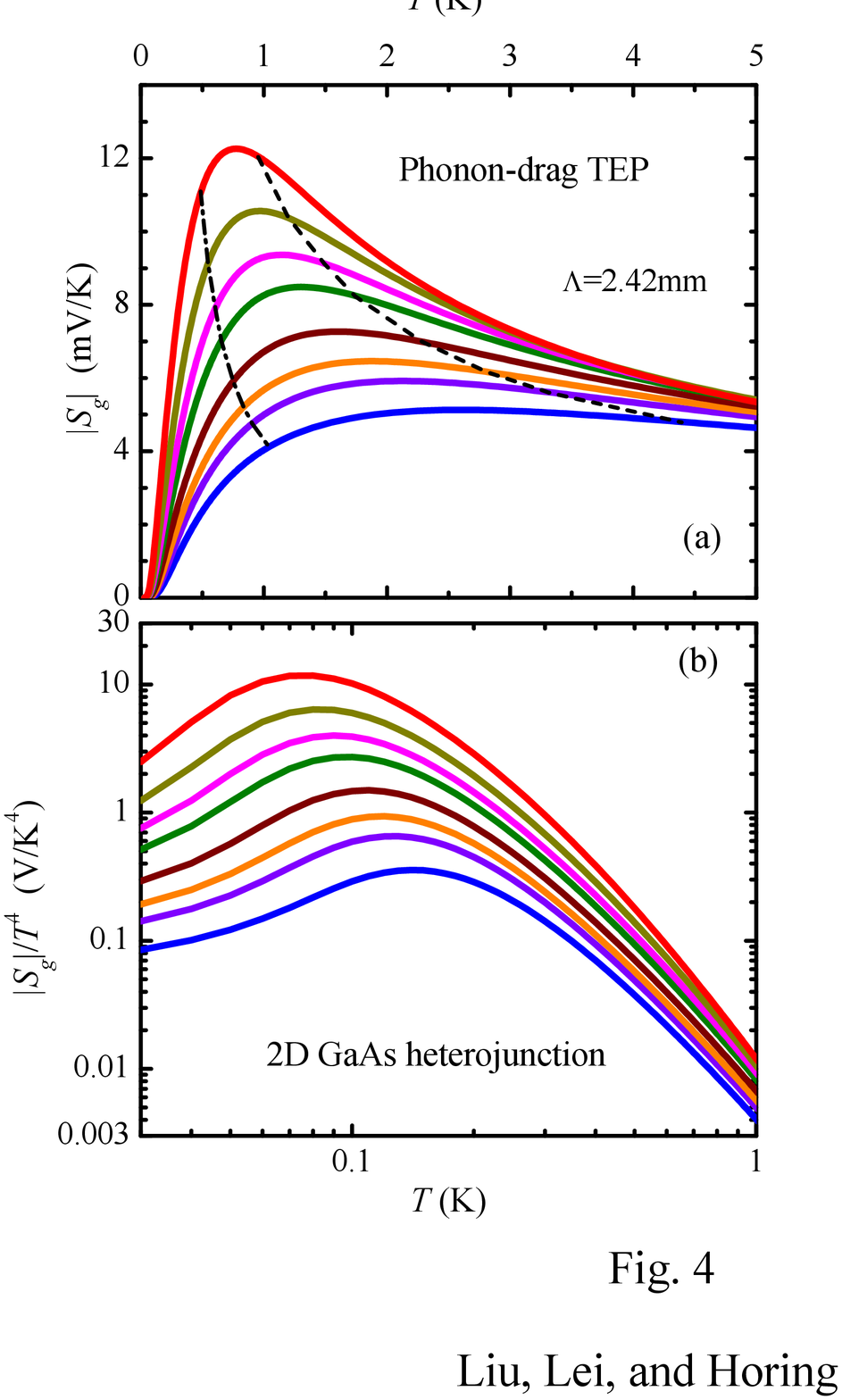}
\caption{(Color online) Temperature dependencies of (a) $|S_g|$
  and (b) $|S_g|/T^4$ for various densities of electrons (from the top): $n_s=0.23$,
  $0.29$, $0.36$, $0.42$, $0.55$, $0.68$,
  $0.80$, and $1.06\times
  10^{10}$\,cm$^{-2}$. The dotted and dashed lines in
  Fig.\,4(a) indicate $T_{\rm BG}$ and
  $\varepsilon_F$ (in Kelvin), respectively.} \label{fig4}
\end{figure}

In Fig.\,4, we plot the temperature dependencies
of phonon-drag thermoelectric power for various
electron densities. We see that $S_g$ vs $T$ for
dilute $n_s$ is significantly different from that in
2D systems having a dense electron density. It is
clear that for relatively
high $n_s$ (for example in the case
$n_s=1.06\times 10^{10}$\,cm$^{-2}$), $|S_g|$
increases as $T$ increases and then it saturates
at a relatively high temperature. This 
can be explained qualitatively by means of the asymptotic
behavior of $S_g$ in high-$n_s$ limit, 
presented in the Appendix: when $T\ll \varepsilon_F$,
$S_g$ first increases with an increase of temperature as
$\sim T^4$ and it becomes independent of
temperature at high temperature. From
Fig.\,4, we also see that for
relatively low $n_s$, a peak appears in $|S_g|$
vs $T$. The position of peak depends on electron
density: the peak moves towards the low temperature
side with a decrease of electron density, but it
lies always between $T_{\rm BG}$ and
$\varepsilon_F$. 

It should be noted that the appearance of a peak in $S_g$
vs $T$ can be understood as the result of competition 
between (i) broadening of the Fermi distribution
function and (ii)
decrease of the rate of nonequilibrium phonon production, induced by
an increase of temperature. As $T$ increases, the Fermi
distribution broadens and dragging electrons out of
equilibrium by nonequilibrium
phonons is facilitated. As a result, the phonon-drag TEP
$|S_g|$ increases with increasing
$T$. However, as $T$ further increases, the rate of
nonequilibrium phonon production induced by a temperature
gradient decreases, leading to a
decrease of $S_g$ with the further increase of
$T$. Competition of these two factors results in
the nonmonotonic dependence of $S_g$ on $T$.

From analysis presented above, it is clear that to
observe the nonmonotonic dependence of
$|S_g|$ on $T$ in the presence of only boundary
scattering in the phonon relaxation process, two
conditions are required. One condition
is that the
Fermi energy should be much smaller than the
critical temperature at which phonon-phonon
scattering is important in phonon-relaxation.
In 2D GaAs systems, such a critical
temperature is about $10$\,K,\cite{Wu1996} leading
to an estimate of electron density in a
2D GaAs system for observation of the peak in $S_g$ vs
$T$ as $n_s\lesssim 2.5\times
10^{10}$\,cm$^{-2}$. The second condition for
observation of the peak in $S_g$ vs
$T$ is that $T_{\rm BG}$ should be comparable with
$\varepsilon_F$. Otherwise, the peak
disappears: as temperature increases, $S_g$
monotonically increases and reaches saturation at a
relatively high temperature when $T_{\rm BG}\ll \varepsilon_F$. 

Note that in typical thermoelectric experiments,
the measurable quantity is the total
thermoelectric power, causing
difficulty to separate the diffusion and
phonon-drag contributions. In the dilute 2D
systems studied here, the phonon-drag TEP is
dominant over almost the whole temperature
regime $T<5$\,K. For example, for
$n_s=1.06\times 10^{10}$\,cm$^{-2}$, $|S_g|$
exceeds $|S_d|$ when $T\gtrsim 0.1$\,K. Hence, in
the temperature dependence of the total TEP
$S=S_d+S_g$, which is plotted in Fig.\,5, the
features are almost the same as those
in $S_g$ vs $T$. Therefore, to observe the
nonlinear dependence of diffusion TEP on $T$ as shown in
Fig.\,2, specific structures, such as a
hot-electron thermocouple,\cite{Chickering2009} are required.
\begin{figure}
\includegraphics [width=0.45\textwidth,clip] {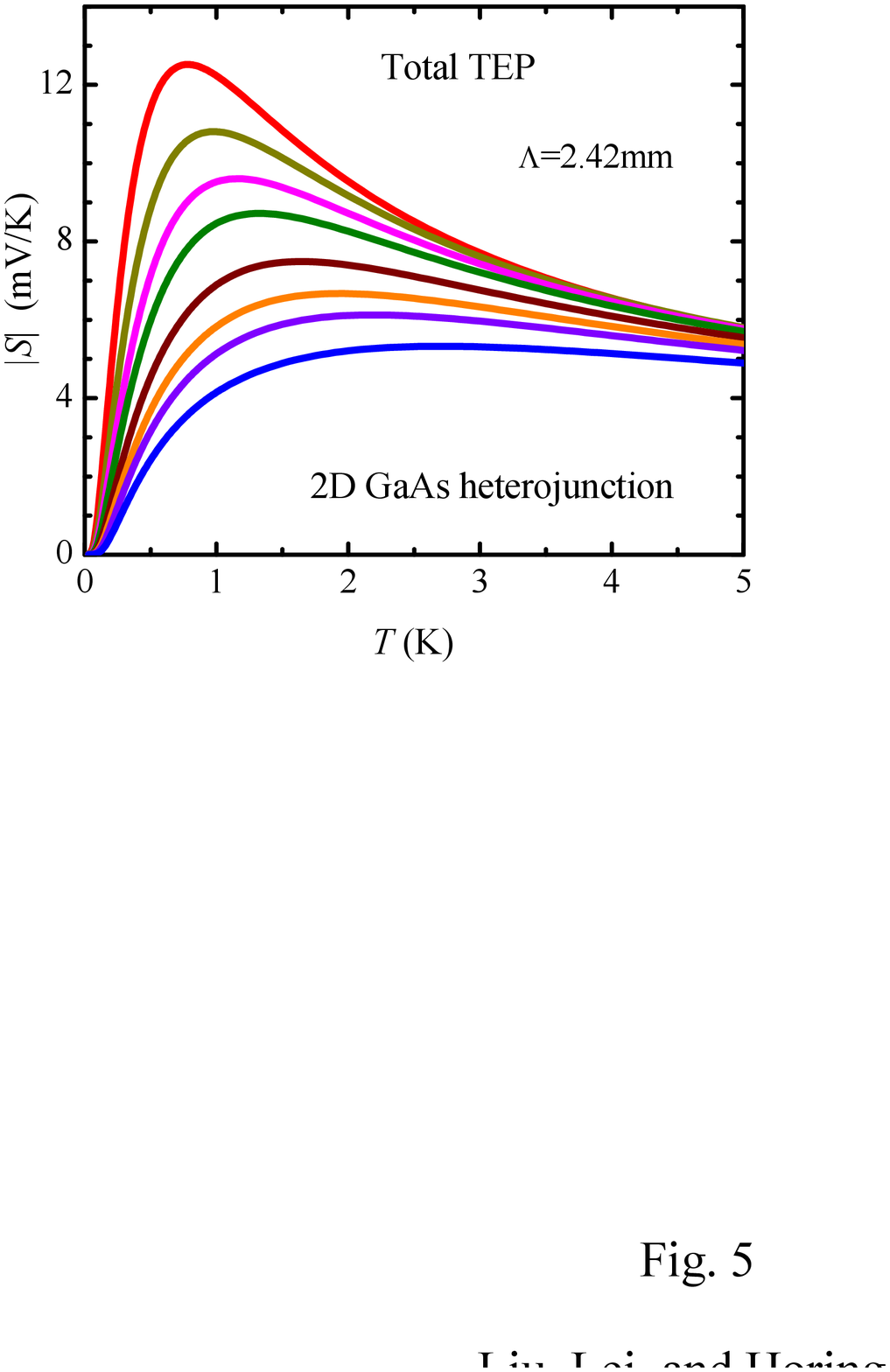}
\caption{(Color online) Temperature dependencies of 
  magnitude of total
  thermoelectric power $S\equiv S_d+S_g$ for various densities of electrons (from the top): $n_s=0.23$,
  $0.29$, $0.36$, $0.42$, $0.55$, $0.68$,
  $0.80$, and $1.06\times
  10^{10}$\,cm$^{-2}$.} \label{fig5}
\end{figure}

\section{Conclusions}

Employing the energy expansion method to solve the
Boltzmann equation and taking account of the screening
of interactions in terms of finite-temperature
STLS theory, we have carried out a theoretical
investigation of the
thermoelectric effect in a two-dimensional electron
system with dilute electron density $0.23\le n_s\le
1.06\times 10^{10}$\,cm$^{-2}$. The temperature dependencies of
the diffusion and phonon-drag thermoelectric powers
have been carefully analyzed for $T\le 5$\,K and
our results exhibit deviations from the conventional
simple power laws. 
We find that, in dilute 2D systems, $|S_d|/T$
remains constant only when $T\lesssim
0.1\varepsilon _F$ and it decreases with an
increase of temperature out of this regime. We
also observe a peak in the temperature
dependence of $|S_g|$, which arises from
competition between thermal broadening of distribution
functions and decrease of the rate of nonequilibrium
phonon production, induced by a temperature increase. 

\begin{acknowledgments}
 This work was supported by the projects of
the National Basic Research Program of
 China (973 Program) (No. 2011CB925603),
the National Science Foundation of China, the National Basic Research
Program of China (2007CB310402), the Shanghai Municipal Commission of Science and Technology (06dj14008), and the
Program for New Century Excellent Talents in University.
\end{acknowledgments}

\appendix*

\section{Asymptotic behaviors of $S_d$ and
  $S_g$ in the high-electron-density limit}

We verify that the energy-expansion method presented
in section II produces the conventional
expressions for $S_d$ and $S_g$ in the case $T\ll \varepsilon_F$
[$\varepsilon_F\equiv 2\pi n_s/(g_sm^*)$ is the
Fermi energy]. Obviously, in high-electron-density
limit, it is sufficient to
consider only the two lowest terms in the expansion of $g_{\bm p}$:
$g_{\bm p}\approx
[C_0\eta_0+C_1\eta_1(\varepsilon)]\cos\varphi_{\bm
  p}$.
Based on the orthonormality conditions of
$\eta_n(\varepsilon)$, Eq.\,(\ref{OR}),
$\eta_0(\varepsilon)$ and
$\eta_1(\varepsilon)$ can be written, respectively, as
\begin{eqnarray}
\eta_0(\varepsilon)&=&\frac
{1}{\sqrt{f_0(0)}}\nonumber\\
&=&1+\frac 12 {\rm e}^{-\frac{\varepsilon_F}{T}}
+\frac{3}{8} {\rm e}^{-\frac{2\varepsilon_F}{T}}
  +O\left ( {\rm e}^{-\frac{3\varepsilon_F}{T}}\right )\label{A1}
    \end{eqnarray}
and
\begin{eqnarray}
\eta_1(\varepsilon)&=&\frac{\sqrt{3}(\varepsilon_F-\varepsilon)}{\pi
  T}+\left \{\frac{3\sqrt{3}(\varepsilon_F-\varepsilon)}{2\pi^3
  T}\right .\nonumber\\
&&\times\left .\left[\frac 12\left
      (\frac{\varepsilon_F}{T}\right)^2+\frac{\varepsilon_F}{T}+1\right
  ]+\frac{\sqrt{3}}{\pi}\frac{\varepsilon_F}{T}\right\}{\rm
  e}^{-\frac{\varepsilon_F}{T}}\nonumber\\
&&  +O\left ( T^{-5}{\rm e}^{-\frac{2\varepsilon_F}{T}}\right ).\label{A2}
    \end{eqnarray}
From Eqs.\,(\ref{A1}) and (\ref{A2}) it is obvious that the
corrections to the leading terms of
$\eta_{0,1}(\varepsilon)$ are exponentially
small for $T\ll\varepsilon_F$ and hence $\eta_0(\varepsilon)=1$ and
$\eta_1(\varepsilon)=\sqrt{3}(\varepsilon_F-\varepsilon)/(\pi
T)$ can be used in the calculation that follows.

\subsection{Temperature dependence of resistivity
  in the high-electron-density limit}

Before proceeding to analyze $S_d$ and $S_g$
{vs} $T$, it is useful to evaluate the
temperature dependence of resistivity, which 
is defined as
$\rho=1/\sigma=m^*\left <\tau^{-1}\right
>/(n_se^2)$ ($\left <\tau^{-1}\right >$ is the average inverse
relaxation time) and is proportional to $Q_{00}$
in high-electron-density limit:
$\rho=\eta_0^3Q_{00}/(g_se^2)=\eta_0^3(Q_{00}^{\rm
  imp}+Q_{00}^{\rm ph})/(g_se^2)$. 

We first
consider the temperature dependence of $Q_{00}^{\rm
  imp}$ resulting from RPA-screened
electron-impurity interaction [$G(q)=0$ is used].
Using the potential given by Eq.\,(\ref{POT}), $Q_{00}^{\rm imp}$ 
  can be expressed as
\begin{equation}
Q_{00}^{\rm imp}=\int d\varepsilon
\delta(\varepsilon_F-\varepsilon)\Lambda^{\rm imp}(\varepsilon),\label{QL}
\end{equation}
with $\Lambda^{\rm imp}(\varepsilon_F)$
taking the form 
\begin{eqnarray}
\Lambda^{\rm
  imp}(\varepsilon_F)&=&\frac{1}{\varepsilon_F^2}\int_0^{2k_F}dq\left
  |\frac{
    U_{q}}{\epsilon(q,0)}\right|^2\frac{q^2}{\sqrt{4k_F^2-q^2}}.
\end{eqnarray}
In the low-temperature limit, $\Lambda^{\rm
  imp}(\varepsilon_F)$ can be further
expanded as
\begin{eqnarray}
\Lambda^{\rm imp}(\varepsilon_F)
&=&\frac{1}{\varepsilon_F^2}\int_0^{2k_F}dq\left
  |\frac{
    U_{q}}{\epsilon_{T=0}(q,0)}\right|^2\frac{q^2}{\sqrt{4k_F^2-q^2}}\nonumber\\
&&+8\sqrt{\frac{T}{2\varepsilon_F^3}}|
U_{2k_F}|^2
\sum_{n=0}^{\infty}\left
  (\frac{m^*}{2\pi}\right)^n\frac{\left(\widetilde V_{2k_F}H_{2k_F}\right)^n}{[\epsilon_{T=0}(2k_F,0)]^{n+2}}
  \nonumber\\
&&\times (n+1)\left(\frac{4m^*T}{k_F^2}\right)^{n/2}
 \int_0^\infty\frac{dx}{\sqrt{x}}\phi(x),\label{DQC}
\end{eqnarray}
with $\phi(x)=\int_0^\infty
dy\frac{\sqrt{y}}{\cosh^2(y+x)}$.\cite{Gold1986}
From Eqs.\,(\ref{QL}) and (\ref{DQC}) it follows
that, for $T\ll \varepsilon_F$,  $Q_{00}^{\rm
  imp}$ can be written as
\begin{equation}
Q_{00}^{\rm
  imp}=\Lambda^{\rm imp}(\varepsilon_F)=
\left. Q_{00}^{\rm
  imp}\right |_{T=0}+\Gamma^{\rm imp}(\varepsilon_F) T+O(T^{3/2}),\label{DQ1}
\end{equation}
with $\Gamma^{\rm imp}(\varepsilon_F)$ 
determined by
\begin{equation}
\Gamma^{\rm imp}(\varepsilon_F)=\frac{8{m^*}}{\pi\varepsilon_F^2}
\frac{\widetilde V_{2k_F}H_{2k_F}}{[\epsilon_{T=0}(2k_F,0)]^{3}}\int_0^\infty\frac{dx}{\sqrt{x}}\phi(x),
\end{equation}
and $\left. Q_{00}^{\rm
  imp}\right |_{T=0}$ is obtained from
Eq.\,(\ref{DQC}) by setting $T=0$.
From Eq.\,(\ref{DQ1}) it is clear that
the first-order finite-temperature correction to $Q_{00}^{\rm imp}$ is linear in
$T$, consistent with previous transport
studies.\cite{Gold1986,DasSarma2004}
Note that such a correction
comes mainly from the temperature dependence of the
dielectric function in the screened electron-impurity
scattering potential. Moreover, if the
screening of electron-impurity scattering is
considered by means of the finite-temperature STLS
theory, an additional temperature dependence
associated with $G(q)$ needs to be taken into account.

Further, considering both the
deformation and piezoelectric scatterings, we
carry out the determination of the temperature dependence of
$Q_{00}^{\rm ph}$ both in the BG ($T\ll
T_{\rm BG}\ll \varepsilon_F$) and in the equipartition
($T_{\rm BG}\ll T\ll \varepsilon_F$) regimes. In both cases,
$T_{\rm BG}\ll\varepsilon_F$ and hence
we can make the approximation: 
\begin{equation}
f_0(\varepsilon_{\bm p})[1-f_0(\varepsilon_{\bm
  p}+\Omega_{\bm Q\lambda})]\approx (1+n_{\bm
  Q\lambda})\Omega_{\bm
  Q\lambda}\delta(\varepsilon_{\bm p}-\varepsilon_F).
\end{equation} 
Performing the $\bm p$-integration in
Eq.\,(\ref{Qph}), $Q_{nn'}^{\rm ph}$
for $n=n'=0$ can be written as
\begin{equation}
Q_{00}^{\rm ph}=\int d\varepsilon
\delta(\varepsilon_F-\varepsilon)\Lambda^{\rm ph}(\varepsilon)
\end{equation}
with $\Lambda^{\rm ph}(\varepsilon)$ defined by
($k_F=\sqrt{2m^*\varepsilon_F}$)
\begin{eqnarray}
\Lambda^{\rm ph}(\varepsilon_F)&\approx&\frac{1}{4\pi
  T\varepsilon_F^2}\sum_{\lambda,\pm}\int_{-\infty}^{\infty}|I(iq_z)|^2dq_z\int_0^\infty
qdq | M_{\bm Q\lambda}|^2\nonumber\\
&&\times
\Omega_{\bm Q\lambda}G_{\pm}(q,\Omega_{\bm Q\lambda})n_{\bm
  Q\lambda}(n_{\bm
  Q\lambda}+1).\label{Qph00}
\end{eqnarray}
Here, $G_{\pm}(q,\Omega_{\bm
  Q\lambda})\equiv 2(\varepsilon_{\bm q}\pm\Omega_{\bm Q\lambda})[(k_Fq/m^*)^2-(\Omega_{\bm
  Q\lambda}\pm \varepsilon_{\bm q})^2]^{-1/2}$
is associated with $\varphi_{\hat {\bm p\bm
    q}}$-integration over $\delta$ function. 

In the BG regime, $T\ll T_{\rm BG}\ll
\varepsilon_F$, $G_\pm(q,\Omega_{\bm Q\lambda})\approx
2m^*(\varepsilon_q\pm \Omega_{\bm Q\lambda})/(k_F q)$, $|I(iq_z)|^2\rightarrow 1$, and
$\epsilon(q,\Omega_{\bm Q\lambda})\approx q_s/q$ with
$q_s=m^*e^2/(2\pi\varepsilon_0\kappa)$ as the
screening wave vector. Thus, Eq.\,(\ref{Qph00}) can be
rewritten in the low-temperature limit as 
\begin{eqnarray}
\Lambda^{\rm ph-BG}(\varepsilon_F)&=&\frac{T^4}{4\pi
  k_F\varepsilon_F^2}\sum_\lambda \frac{1}{u_{s\lambda}^4}\int_{-\infty}^\infty d\bar q_z
\nonumber\\
&&\times\int_0^\infty
d\bar q | M_{\bm Q\lambda}|^2\frac{\bar q^2\bar Q {\rm e}^{\bar
    Q}}{\left ({\rm e}^{\bar Q}-1\right )^2},\label{Qph100}
\end{eqnarray}
with $\bar Q=\sqrt{\bar q_z^2+\bar q^2}$.
Substituting the explicit form of $ M_{\bm
  Q\lambda}$ into Eq.\,(\ref{Qph100}),
the contribution to $Q_{00}^{\rm ph}$ in the BG regime
from the deformation potential, $Q_{00}^{\rm DP-BG}$,
is given by
\begin{eqnarray}
Q_{00}^{\rm DP-BG}&=&\frac{T^5\Xi^2}{8\pi d
  k_F\varepsilon_F^2u_{sl}^6}
\int_{-\infty}^\infty d\bar q_z\int_0^\infty
d\bar q \frac{\bar q^2\bar Q^2 {\rm e}^{\bar
    Q}}{\left ({\rm e}^{\bar Q}-1\right
  )^2}\nonumber\\
&=&\frac{15\zeta(5)T^5\Xi^2}{ 2d
  k_F\varepsilon_F^2u_{sl}^6},\nonumber\\ \label{QDP}
\end{eqnarray}
and the contribution from the screened
piezoelectric interaction to
$Q_{00}^{\rm ph-BG}$, $Q_{00}^{\rm PZ-BG}$, can be
written as [$\bar B_{\rm LA}\equiv9\bar q^4\bar q_z^2/\bar Q^7$ and
$\bar B_{\rm TA}\equiv(\bar q^6+8\bar q^2\bar q_z^4)/\bar Q^7$]
\begin{eqnarray}
Q_{00}^{\rm PZ-BG}&=&\frac{\pi T^5 e^2e_{14}^2}{\kappa^2 d
  k_F\varepsilon_F^2q_s^2}\sum_{\lambda={\rm LA,TA}}\frac{1}{u_{{
  s} \lambda}^6}\nonumber\\
&&\times
\int_{-\infty}^\infty d\bar q_z\int_0^\infty
d\bar q \frac{\bar B_\lambda\bar q^4\bar Q {\rm e}^{\bar
    Q}}{\left ({\rm e}^{\bar Q}-1\right
  )^2}\nonumber\\
&=&\frac{45\pi^2\zeta(5) T^5 e^2e_{14}^2}{32\kappa^2 d
  k_F\varepsilon_F^2q_s^2}\left (\frac{21}{ u_{sl}^6}+\frac{29}{ u_{st}^6}\right ),\label{QPZ}
\end{eqnarray}
with $\zeta(x)$ as the Riemann function:
$\zeta(5)\approx 1.037$. 
To
derive Eq.\,(\ref{QPZ}), the $\varphi_{\bm
  q}$-independent forms of the scattering matrix
are used: $|M_{\bm Q\lambda}|^2_{\rm
  piez}=4\pi^2e^2e_{14}^2B_{\lambda}/\left
  [\kappa^2d u_{s\lambda}|\varepsilon(\bm
q,\Omega_{\bm Q\lambda})|^2\right ]$ [$ B_{\rm
LA}\equiv9 q^4 q_z^2/ Q^7$ and
$B_{\rm TA}\equiv(q^6+8 q^2 q_z^4)/ Q^7$],\cite{Lyo1988}
which differ slightly from those presented in
Sec. III. 

Using the material
parameters for a GaAs/AlGaAs heterojunction,
$Q_{00}^{\rm DP-BG}$ and $Q_{00}^{\rm PZ-BG}$ vary with temperature as
$Q_{00}^{\rm DP-BG}\approx 1.936\times
10^{-5}T^5/n_s^{3/2}$ and $Q_{00}^{\rm PZ-BG}\approx 3.085\times
10^{-4}T^5/n_s^{3/2}$ ($n_s$ in
$10^{11}$\,cm$^{-2}$), respectively. This implies
that, of the various
electron-phonon scatterings, the (longitudinal-phonon)
piezoelectric interaction is dominant at low temperature. 

In the EP regime,  $T_{\rm BG}\ll T\ll \varepsilon_F$
and $n_{\bm
  Q\lambda}\approx n_{\bm
  Q\lambda}+1\approx T/\Omega_{\bm Q\lambda}
$, and Eq.\,(\ref{Qph00}) reduces to
\begin{eqnarray}
\Lambda^{\rm ph-EP}(\varepsilon_F)&\approx&\frac{T}{4\pi
  k_F\varepsilon_F^2}\sum_{\lambda}\int_{-\infty}^{\infty}|I(iq_z)|^2dq_z\int_0^{2k_F}
dq\nonumber\\
&&\times | M_{\bm Q\lambda}|^2\frac{q^2/\Omega_{\bm q\lambda}}{\sqrt{1-\frac{q^2}{4p^2}}}.\label{Qph1s}
\end{eqnarray}
Accordingly, the contribution to $Q_{00}^{\rm ph-EP}$
from deformation and piezoelectric scatterings in
the EP
regime, $Q_{00}^{\rm DP-EP}$ and $Q_{00}^{\rm
  DP-EP}$, are given by
\begin{eqnarray}
Q_{00}^{\rm DP-EP}&=&\frac{\Xi^2Tm^*}{2 d
  \varepsilon_Fu_{ sl}^2}\int_{-\infty}^{\infty}|I(iq_z)|^2dq_z,\label{Qph12DP}
\end{eqnarray}
and
\begin{eqnarray}
Q_{00}^{\rm PZ-EP}&=&\frac{32\pi T m^*e^2 e_{14}^2}{
  \kappa^2d\varepsilon_Fk_F}\sum_{\lambda}\frac{1}{u_{
    s\lambda}^2}\int_{-\infty}^{\infty}|I(2ik_F\bar
q_z)|^2d\bar q_z\nonumber \\
&&\times\int_0^1d\bar
q\frac{\bar q^2\bar B_\lambda/\bar Q}{[1+q_s/(2k_F\bar
  q)]^2\sqrt{1-\bar q^2}}.\label{Qph12PZ}
\end{eqnarray}
Thus, we find that the
electron-acoustic-phonon scattering
tends to $\rho\sim Q_{00}^{\rm ph}\sim T^5$ in the
BG regime
and $\rho\sim Q_{00}^{\rm ph}\sim T$ in the EP regime.

It should be noted that the resistivity correction
we found here in the analysis of electron-impurity
scattering, which goes beyond the earlier
``linear-in-$T$'' result, is consistent with
previous transport studies in
Refs.\,\onlinecite{Gold1986,DasSarma2004}. In
regard to electron-phonon scattering, our result
concerning the power-law temperature dependence of resistivity
due to piezoelectric scattering agrees with the previous
one: in Refs. \onlinecite{Price1984,Stormer1990}
the contribution to
inverse relaxation time due to piezoelectric scattering
was found to be proportional to $T^5$.
However, our deformation-scattering result is
different from the $T^7$ law obtained previously.
This is associated with the fact that 
the deformation scattering is taken to be
unscreened in present paper, while a screened one was used in
the previous studies.\cite{Price1984,Stormer1990}

\subsection{Diffusion thermoelectric power in the high-electron-density limit}
To obtain the diffusion TEP $S_d$, one needs to consider the second
term on the left hand side of Eq.\,(\ref{BE_R}). This term can be written for $n=0$ as
\begin{eqnarray}
\frac{\nabla_x T}{T}\sum_{m}\alpha_{0m}\gamma_m&=&
-\nabla_x T\left(1-\frac{\varepsilon_F}{T}\right ){\rm
  e}^{-\frac{\varepsilon_F}{T}}\nonumber \\
&&-\nabla_x T\frac{\varepsilon_F}{2T}{\rm
  e}^{-\frac{2\varepsilon_F}{T}}+O\left ({\rm
    e}^{-\frac{3\varepsilon_F}{T}}\right ),\nonumber\\
\end{eqnarray}
and for $n=1$ we have
\begin{eqnarray}
\frac{\nabla_x
  T}{T}\sum_{m}\alpha_{1m}\gamma_m&=&\frac{\sqrt{3}\pi}{3}\nabla_x
T-\nabla_x
T\frac{\sqrt{3}}{\pi}{\rm
  e}^{-\frac{\varepsilon_F}{T}}\nonumber\\
&&\times\left [1+\frac{\varepsilon_F}{T}+\left
    (\frac{\varepsilon_F}{T}\right)^2\right ]+O\left ({\rm
    e}^{-\frac{2\varepsilon_F}{T}}\right ).\nonumber\\
\end{eqnarray}
It is clear that, in high-electron-density limit, the $n=0$ term is exponentially
small while the term with $n=1$ is dominant and
is
given by
\begin{equation}
\frac{\nabla_x
  T}{T}\sum_{m}\alpha_{1m}\gamma_m\approx\frac{\sqrt{3}\pi}{3}\nabla_x
T.
\end{equation}

Assuming
$\eta_n(\varepsilon_{\bm p})\approx
\eta_n(\varepsilon_{\bm p-\bm q})$  in the case $T,T_{\rm BG}\ll
\varepsilon_F$, $Q_{nn'}^{\rm
  imp.ph}$ can be written as
($n,n'\le 1$)
\begin{eqnarray}
Q_{nn'}^{\rm
  imp,ph}&=&\int_0^\infty d\varepsilon \left
  [-\frac{\partial f_0(\varepsilon)}{\partial
    \varepsilon}\right
]\nonumber\\
&&\times\left
  [\frac{\sqrt{3}(\varepsilon_F-\varepsilon)}{\pi T}\right]^{n+n'}\Lambda^{\rm imp,ph}(\varepsilon).
\end{eqnarray}
Using the low-temperature expansion of the Fermi
function\cite{VKLukyanov1995}
\begin{equation}
f_0(\varepsilon)=\Theta(\varepsilon_F-\varepsilon)-\frac{\pi^2}{6}T^2\delta^{(1)}(\varepsilon-\varepsilon_F)-\frac{7\pi^4}{30}T^4\delta^{(3)}(\varepsilon-\varepsilon_F)+...
\end{equation}
and performing the energy integration,
the leading terms of $Q_{nn'}^{\rm imp,ph}$ take
the forms
\begin{eqnarray}
Q_{10}^{\rm imp,ph}=Q_{01}^{\rm
  imp,ph}=-\frac{\sqrt{3}\pi T}{3}\frac{\partial
  \Lambda^{\rm imp,ph}(\varepsilon_F)}{\partial \varepsilon_F},\label{Qimp10}
\end{eqnarray}
and 
\begin{eqnarray}
Q_{11}^{\rm imp,ph}=Q_{00}^{\rm imp,ph}=\Lambda^{\rm imp,ph}(\varepsilon_F).\label{Qimp11}
\end{eqnarray}
Substituting these $Q_{nn'}$ terms into
Eq.\,(\ref{BE_R}), the solution
$C_0^{({\rm d})}$ can be written as
\begin{equation}
C_0^{({\rm d})}\approx
\nabla_x T\frac{\pi^2}{3}T\frac{\partial
}{\partial \varepsilon_F}\left [\Lambda^{\rm
    imp}(\varepsilon_F)+\Lambda^{\rm
    ph}(\varepsilon_F)\right]^{-1},
\end{equation}
and the diffusion TEP takes the form 
\begin{eqnarray}
S_{
  d}&\approx&-\frac{\pi^2T}{3e}\left. \Lambda^{\rm
    imp}(\varepsilon_F)\right
|_{T=0}\frac{\partial }{\partial
  \varepsilon_F}\frac{1}{\left .\Lambda^{\rm
    imp}(\varepsilon_F)\right |_{T=0}}\nonumber\\
&&+\frac{\pi^2T}{3e}\frac{\partial}{\partial
  \varepsilon_F}\left [\frac{\Gamma(\varepsilon_F)T+\Lambda^{\rm
    ph}(\varepsilon_F)}{\left .\Lambda^{\rm
    imp}(\varepsilon_F)\right |_{T=0}}\right
].
\label{Mott}
\end{eqnarray}

From Eq.\,(\ref{Mott}) we see that the first term
on the right hand side agrees with  the
well-known Mott formula.\cite{Mott1979}
Considering the fact that
$Q_{00}$ relates to $n_s$ approximately as
$Q_{00}\approx m^*\left
  <\tau^{-1}\right >/(g_sn_s)$
and ignoring the
energy-dependence of $\left <\tau^{-1}\right >$, $S_d\approx
\pi^2T/(3e\varepsilon_F)$ can be
obtained. However, if one assumes $\left <\tau^{-1}\right >\sim
\varepsilon_F^p$,  we obtain $S_d\approx
\pi^2T/(3e\varepsilon_F)(1+p)$, in agreement with the
results of Refs.\onlinecite{VKKaravolasetal1990,VCKaravolasandPNButcher1991}.  

In Eq.\,(\ref{Mott}), the second term on
the right hand side is a low-temperature correction to the leading
term and it is proportional to $T^2$. Obviously,
in the BG regime, this
correction comes only from the
temperature dependence of the screening of
electron-impurity scattering, since the phonon
contribution is proportional to $T^6$ in
BG regime [$\Lambda^{\rm
  ph}(\varepsilon_F)\sim T^5$] and it can be
ignored. However, in the
equipartition regime, the electron-phonon
scattering results in $\Lambda^{\rm
    ph}(\varepsilon_F)$ being linear in $T$. Hence, both
  the electron-impurity and electron-phonon
  scatterings lead to a deviation of $S_d$ vs $T$
  from the
  linear rule when $T_{\rm BG}\ll T\ll \varepsilon_F$. 

\subsection{Phonon-drag thermoelectric power in
  the high-electron-density limit}

To investigate the phonon-drag effect in
thermoelectric power, one needs to study the
driving term $D_n$ in
Eq.\,(\ref{BE_R}). Performing substitution, $\bm
q\rightarrow -\bm q$ for $+$ term and
$\bm p\rightarrow \bm p+\bm q$ for $-$ term,
Eq.\,(\ref{D}) can be rewritten as
\begin{eqnarray}
D_n&=&2\pi\sum_{{\bm Q},{\bm
    p},\lambda}|\widetilde M_{{\bm Q
    }\lambda}|^2\delta(\varepsilon_{{\bm
    p}+{\bm q}}-\varepsilon_{{\bm p}}-\Omega_{{\bm
    Q}\lambda})\tau_{{\rm p}\lambda}u^x_{\bm
  Q\lambda} \nonumber\\
&&\times \frac{\Omega_{\bm Q
    \lambda}}{T}n_{{\bm Q}\lambda}(1+n_{{\bm
    Q}\lambda})\left [f_0(\varepsilon_{\bm p})-f_0(\varepsilon_{{\bm p}+{\bm
      q}})\right ]\nonumber\\
&&\times\left [\chi_n(\bm p +\bm q)-\chi_n(\bm
  p)\right ].\label{DD}
\end{eqnarray}

Considering
only the driving term $D_n$ with $n=0,1$, the
solution of Eq.\,(\ref{BE_R}), $C_0^{(\rm g)}$, can be
written as
\begin{eqnarray}
C_0^{(\rm
  g)}&=&\frac{\nabla_xT}{T}\frac{Q_{11}D_0-Q_{01}D_1}{Q_{11}Q_{00}-Q_{01}Q_{10}}\nonumber\\
&\approx&
\frac{\nabla_xT}{T}\left (\frac 1{Q_{00}}
  D_0-\frac{Q_{01}}{Q_{00}Q_{11}}D_1\right ).
\end{eqnarray} 
Using Eqs.\,(\ref{Qimp10}) and (\ref{Qimp11}), the
phonon-drag $S_g$ is determined by
\begin{equation}
S_g=-\frac{1}{eT}\left [D_0+\frac{\sqrt{3}}{3}\pi
  T D_1\frac{\partial }{\partial
    \varepsilon_F}\ln \Lambda(\varepsilon_F)
\right ].
\end{equation}
Recognizing that $\chi_n(\bm p+\bm q)-\chi_n(\bm p)\approx
4\pi q\cos\varphi_{\bm q}\eta_n(\varepsilon_{\bm
  p})/p^2$ in the case $T\ll \varepsilon_F$, $S_g$ finally 
takes the form 
\begin{eqnarray}
S_g&=&-\frac{2\pi^2}{m^*eT}\frac{\partial \ln
  \Lambda(\varepsilon_F)}{\partial
    \varepsilon_F}
  \sum_{{\bm Q},{\bm
    p},\lambda}|\widetilde M_{{\bm Q
    }\lambda}|^2\delta(\varepsilon_{{\bm
    p}+{\bm q}}-\varepsilon_{{\bm p}}-\Omega_{{\bm
    Q}\lambda}) \nonumber\\
&&\times \tau_{{\rm p}\lambda}u_{\bm
  Q\lambda}\frac{q^2\Omega_{\bm Q
    \lambda}}{QT}n_{{\bm Q}\lambda}(1+n_{{\bm
    Q}\lambda})\left [f_0(\varepsilon_{\bm p})-f_0(\varepsilon_{{\bm p}+{\bm
      q}})\right ].\label{S_g_HT}
\end{eqnarray}
Note that this expression for $S_g$ reduces to the
one widely used in literature\cite{Cantrell1987,Khveshchenko1997,Fletcher2002} if $\frac{\partial \ln
  \Lambda(\varepsilon_F)}{\partial
    \varepsilon_F}$ is replaced by
  $1/\varepsilon_F$.

To further analyze the power law of $S_g$ {vs}
$T$ in the high-$n_s$ limit, one has to study the temperature dependence
of $\tau_{{\rm p}\lambda}$.
 At sufficiently low temperature, it is reasonable to
 assume that boundary scattering dominates
 phonon relaxation and the mean free path of
 phonons, $\Lambda$, is independent of $T$. Under
 this consideration, $S_g$ in the BG regime,
 $S_g^{\rm BG}$, can be
 written as
\begin{eqnarray}
S_g^{\rm
  BG}&\approx&-\frac{{m^*}^2T^3}{4\pi^2k_Fn_se}\sum_{\lambda}\int_{-\infty}^{\infty}d\bar
q_z\int_0^{\infty}d\bar q\nonumber \\
&&\times\left |M_{\bm Q\lambda}\right
|^2\frac{\tau_{{\rm p}\lambda}\bar q^2 \bar Q
  {\rm e}^{\bar Q}}{u_{s\lambda}^2({\rm e}^{\bar Q}-1)^2},\label{AEND_BG} 
\end{eqnarray}
and, in the EP regime, it takes the form
\begin{eqnarray}
S_g^{\rm
  EP}&\approx&-\frac{{m^*}^2}{4\pi^2k_Fn_se}\sum_{\lambda}\tau_{{\rm
  p}\lambda}u_{s\lambda}\int_{-\infty}^{\infty}d
q_z|I(iq_z)|^2\int_0^{\infty}d q\nonumber \\
&&\times\left |M_{\bm Q\lambda}\right
|^2\frac{q^2}{Q\sqrt{1-\left (\frac{m^*}{pq}\right
    )^2\left (\Omega_{\bm
        Q\lambda}-\varepsilon_{\bm q}\right )^2}}. \label{AEND_EP}
\end{eqnarray}
$S_g^{\rm BG}$ can be further simplified by
substituting explicit forms of the deformation and
piezoelectric scattering matrices into it and then
performing momentum integration:
\begin{eqnarray}
S^{\rm
  BG}_g&\approx&-\frac{15{m^*}^2\Xi^2T^4\tau_{\rm
    p,LA}\zeta(5)}{2\pi\sqrt{2\pi}dn_s^{3/2}eu_{sl}^4}\nonumber\\
&&-\frac{45{m^*}^2e
  e_{14}^2T^4\zeta(5)}{32\pi\sqrt{2\pi}d\kappa^2q_s^2n_s^{3/2}}\left
  (21\frac{\tau_{\rm
    p,LA}}{u_{sl}^4}+29\frac{\tau_{\rm
    p,TA}}{u_{st}^4}\right ).
\end{eqnarray}

From Eqs.(\ref{AEND_BG}) and (\ref{AEND_EP}) we
see that, when
$T\rightarrow 0$, the phonon-drag
thermoelectric power tends to zero as $T^4$ in the
BG
regime and it reaches a saturation value in the EP
regime. Note that such behavior of $S_g$ vs $T$ in
the BG regime has
already been demonstrated in
Refs.\onlinecite{Tieke1998,Khveshchenko1997},
while, as far as we know, the
temperature-independence of $S_g$ in the EP regime,
obtained here, is a new prediction.

\bibliography{Seebeck_Effect}

\end{document}